\documentclass{aa}
\usepackage{graphicx}
\usepackage{hyperref}
\hypersetup{
    colorlinks,
    citecolor=blue,
    filecolor=blue,
    linkcolor=blue,
    urlcolor=blue}
\usepackage{textcomp}
\usepackage{txfonts}
\begin{document}

\newcommand{\af}{A_{\rm F}}
\newcommand{\afphi}{A_{\rm F,PHI}}
\newcommand{\afhmi}{A_{\rm F,HMI}}
\newcommand{\afratio}{\frac{\afhmi}{\afphi}}

\newcommand{\as}{A_{\rm S}}
\newcommand{\asphi}{A_{\rm S,PHI}}
\newcommand{\ashmi}{A_{\rm S,HMI}}
\newcommand{\asratio}{\frac{\ashmi}{\asphi}}

\newcommand{\blos}{B_{\rm LOS}}
\newcommand{\blosabs}{\left|B_{\rm LOS}\right|}
\newcommand{\blosmu}{\blos/\mu}
\newcommand{\blosmusat}{\left(\blosmu\right)_{\rm sat}}

\newcommand{\fb}{\rm FB}
\newcommand{\fbdc}{\fb_{\rm DC}}
\newcommand{\fbrm}{\fb-\fbdc}
\newcommand{\fbphi}{\fb_{\rm PHI}}
\newcommand{\fbhmi}{\fb_{\rm HMI}}

\newcommand{\rn}{\frac{r}{r_{\odot}}}

\newcommand{\sd}{\rm SD}
\newcommand{\sddc}{\sd_{\rm DC}}
\newcommand{\sdrm}{\sd-\sddc}
\newcommand{\sdphi}{\sd_{\rm PHI}}
\newcommand{\sdhmi}{\sd_{\rm HMI}}
\newcommand{\sddcphi}{\sd_{\rm DC,PHI}}
\newcommand{\sddchmi}{\sd_{\rm DC,HMI}}

\title{{Reconstruction of total solar irradiance variability as simultaneously apparent  from Solar Orbiter and Solar
Dynamics Observatory}}
\titlerunning{Reconstruction of TSI variability as seen from Solar Orbiter and SDO}

\author{K.L.~Yeo\inst{1}\thanks{\hbox{Corresponding author: K.L.~Yeo} \hbox{\email{yeo@mps.mpg.de}}}
\and
   N.A.~Krivova\inst{1} \and
   S.K.~Solanki\inst{1} \and
   J.~Hirzberger\inst{1} \and 
   D.~Orozco~Su\' arez\inst{2} \and 
   K.~Albert\inst{1} \and 
   N. Albelo~Jorge\inst{1} \and 
   T.~Appourchaux\inst{3} \and 
   A.~Alvarez-Herrero\inst{4} \and 
   J.~Blanco Rodr\'\i guez\inst{5} \and 
   A.~Gandorfer\inst{1} \and 
   P.~Gutierrez-Marques\inst{1} \and 
   F. Kahil\inst{1} \and 
   M.~Kolleck\inst{1} \and 
   J.C.~del~Toro~Iniesta\inst{2} 
   R.~Volkmer\inst{6} 
   J.~Woch\inst{1} \and 
   B.~Fiethe\inst{7} \and 
   I.~P\' erez-Grande\inst{9} \and 
   E.~Sanchis~Kilders\inst{5} \and 
   M.~Balaguer~Jiménez\inst{2}~\and 
   L.R.~Bellot~Rubio\inst{2} \and 
   D.~Calchetti\inst{1} \and 
   M.~Carmona\inst{8} \and 
   A.~Feller\inst{1} \and 
   G.~Fernandez-Rico\inst{1,9} 
   A.~Fern\' andez-Medina\inst{4} \and 
   P.~Garc\'\i a~Parejo\inst{4} \and 
   J.L.~Gasent~Blesa\inst{5} \and 
   L.~Gizon\inst{1,10} 
   B.~Grauf\inst{1} \and 
   K.~Heerlein\inst{1} \and 
   A.~Korpi-Lagg\inst{1} \and 
   T.~Maue\inst{6,11} \and 
   R.~Meller\inst{1} \and 
   A.~Moreno Vacas\inst{2} \and 
   R.~M\" uller\inst{1} \and
   E.~Nakai\inst{6} \and 
   W.~Schmidt\inst{6} \and 
   J.~Schou\inst{1} \and 
   J.~Sinjan \inst{1} \and 
   J.~Staub\inst{1} \and 
   H.~Strecker \inst{2} \and 
   I.~Torralbo\inst{9} \and 
   G.~Valori\inst{1} 
   }
     
   \institute{
         Max-Planck-Institut f\"ur Sonnensystemforschung, Justus-von-Liebig-Weg 3,
         37077 G\"ottingen, Germany \\ \email{solanki@mps.mpg.de}
         \and
         Instituto de Astrofísica de Andalucía (IAA-CSIC), Apartado de Correos 3004,
         E-18080 Granada, Spain
         \and
         Univ. Paris-Sud, Institut d’Astrophysique Spatiale, UMR 8617,
         CNRS, B\^ atiment 121, 91405 Orsay Cedex, France
         \and
         Instituto Nacional de T\' ecnica Aeroespacial, Carretera de
         Ajalvir, km 4, E-28850 Torrej\' on de Ardoz, Spain
         \and
         Universitat de Val\`encia, Catedr\'atico Jos\'e Beltr\'an 2, E-46980 Paterna-Valencia, Spain
         \and
         Leibniz-Institut für Sonnenphysik, Sch\" oneckstr. 6, 79104 Freiburg, Germany
         \and
         Institut f\"ur Datentechnik und Kommunikationsnetze der TU
         Braunschweig, Hans-Sommer-Str. 66, 38106 Braunschweig,
         Germany
         \and
         University of Barcelona, Department of Electronics, Carrer de Mart\'\i\ i Franqu\`es, 1 - 11, 08028 Barcelona, Spain
         \and
         Instituto Universitario "Ignacio da Riva", Universidad Polit\'ecnica de Madrid, IDR/UPM, Plaza Cardenal Cisneros 3, E-28040 Madrid, Spain
         \and
         Institut f\"ur Astrophysik, Georg-August-Universit\"at G\"ottingen, Friedrich-Hund-Platz 1, 37077 G\"ottingen, Germany
         \and
         Fraunhofer-Institut f\"ur Kurzzeitdynamik, Ernst-Mach-Institut, EMI
         Ernst-Zermelo-Str. 4, 79104 Freiburg, Germany}

\abstract
{{Solar irradiance variability has been monitored almost exclusively from the Earth's perspective. {We present a method to combine the unprecedented observations of the photospheric magnetic field and continuum intensity  from outside the Sun-Earth line, which is being recorded by the Polarimetric and Helioseismic Imager on board the Solar Orbiter mission (SO/PHI), with solar observations recorded from the Earth's perspective to examine the solar irradiance variability from both perspectives simultaneously.} Taking SO/PHI magnetograms and continuum intensity images from the cruise phase of the Solar Orbiter mission and concurrent observations from the Helioseismic and Magnetic Imager onboard the Solar Dynamics Observatory (SDO/HMI) as input into the SATIRE-S model, we successfully reconstructed the total solar irradiance variability as apparent from both perspectives. In later stages of the SO mission, the orbital plane will tilt in such a way as to bring the spacecraft away from the ecliptic to heliographic latitudes of up to $33^{\circ}$. The current study sets the template for the reconstruction of solar irradiance variability as seen from outside the ecliptic from data that SO/PHI is expected to collect from such positions. {Such a reconstruction will be beneficial to factoring inclination into how the brightness variations of the Sun compare to those of other cool stars, whose rotation axes are randomly inclined.}}}
\maketitle

\section{Introduction}
\label{introduction}

Solar irradiance variability is a key natural driver of climate change \citep{gray10,solanki13} and serves as a template for understanding the brightness variability of other cool stars \citep{foukal93,radick18}. {The variability at timescales of days to the solar cycle is established to be mainly driven by photospheric magnetism \citep{shapiro17,yeo17}.}

The photospheric magnetic field is partly confined in kilogauss-strength concentrations, which are  manifested as bright faculae and network (hereinafter referred to collectively as faculae) and as dark sunspots \citep{spruit83}. Solar irradiance fluctuates as the time evolution of the photospheric magnetic field and solar rotation change the prevalence and spatial distribution of bright faculae and dark sunspots on the solar disc. {The intensity excess produced by an individual facular feature varies with time, as its magnetic field emerges, decays, coalesces, or is cancelled out by nearby magnetic flux (time evolution). Likewise, the intensity deficit produced by a given sunspot is modulated by the time evolution of its magnetic field.} At the same time, the intensity excess or deficit produced by the feature is modulated by the combination of solar rotation moving it across the solar disc and the variation in its intensity contrast and projected area with distance from disc centre (rotational modulation).

{The photospheric magnetic field is spatially heterogeneous such that the magnetism on the solar disc, and consequently solar irradiance variability, can appear to be different to observers looking at the Sun from different directions. How solar irradiance variability might change with the heliographic latitude of the observer is of particular interest as it is relevant to how the brightness variations of the Sun compare to those of other cool stars, which come in all inclinations.} Unfortunately, solar irradiance has so far only been monitored from Earth-orbiting satellites \citep{ermolli13,kopp14} {and from the Mars-orbiting MAVEN satellite \citep{jakosky15}.} Although models have been developed to reconstruct solar irradiance variability from observations related to solar magnetism such as activity indices and full solar disc imagery \citep[see reviews by][]{domingo09,yeo14b,chatzistergos22}, before the Solar Orbiter mission \citep{muller20}, no suitable data  had been recorded from outside the Sun-Earth line. {Consequently, attempts to determine how solar variability might appear from outside the ecliptic have no alternatives but to extrapolate this information from models of the Sun or solar observations recorded from the ecliptic \citep{schatten93,knaack01,vieira12,shapiro14,nemec20a,nemec20b}.}

The Solar Orbiter mission, launched in February 2020, is set to change this state of affairs.  The spacecraft is orbiting the Sun with a period of about 180 days such that, over the course of each orbit, it {assumes} a range of positions outside the Sun-Earth line. {At present, the orbital plane is close to the ecliptic. In the later stages of the Solar Orbiter mission, the orbital plane will tilt to bring the spacecraft away from the ecliptic to heliographic latitudes of up to $33^{\circ}$.} While there is no solar irradiance monitor on board, the payload includes the Polarimetric and Helioseismic Imager \citep[SO/PHI,][]{solanki20}, which returns, amongst other data products, full-disc magnetograms and continuum images suitable for reconstructing solar irradiance variability. {The out-of-ecliptic observations SO/PHI is expected to return will present the unique opportunity to reconstruct solar irradiance variability as apparent from outside the ecliptic from actual observations recorded from such positions. Such a reconstruction is missing from existing attempts to factor inclination into how the brightness variations of the Sun compare to those of other cool stars \citep{schatten93,knaack01,vieira12,shapiro14,nemec20a,nemec20b}.}

{We present a method to augment the unprecedented observations SO/PHI is recording from outside the Sun-Earth line with solar observations recorded from the Earth’s perspective in order to examine solar irradiance variability from both perspectives. This is a preliminary study aimed at setting the template for the reconstruction of solar irradiance variability from the out-of-ecliptic observations expected from SO/PHI in the later stages of the Solar Orbiter mission.} Taking SO/PHI full-disc magnetograms and continuum images recorded during the cruise phase of the mission and concurrent observations from the Helioseismic and Magnetic Imager onboard the Earth-orbiting Solar Dynamics Observatory \citep[SDO/HMI,][]{scherrer12}, we reconstruct {total solar irradiance (TSI)} variability as {apparent simultaneously} from both perspectives. For this purpose, we made use of the Spectral And Total Irradiance REconstructions for the Satellite era model \citep[SATIRE-S,][]{fligge00,krivova03,yeo14a,yeo14b,yeo15}, an established model of solar irradiance variability that has been successfully applied to such data from various solar telescopes, including SDO/HMI, to reconstruct this quantity. {In the following, we describe the data set (Sect. \ref{dataselectionreduction}) and the SATIRE-S model (Sect. \ref{satiresmodel}) and we describe how it is adapted here to the purposes of the current study (Sect. \ref{tsireconstruction}), and the results  (Sect. \ref{section3}).} Finally, we give our concluding remarks (Sect. \ref{conclusion}).

\section{Model}
\label{section2}

The SATIRE-S model computes the corresponding total and spectral solar irradiance from a given full-disc {line-of-sight} magnetogram
and the simultaneous continuum intensity image, hereinafter referred to as
an image pair. For this preliminary study, we confine the scope to total solar irradiance. We input concurrent SO/PHI and SDO/HMI image pairs into the model to recover TSI variability as seen from both perspectives.

\subsection{Data selection and reduction}
\label{dataselectionreduction}

\begin{figure}
\centering
\includegraphics{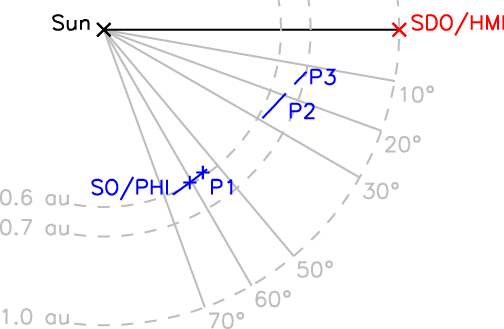}
\caption{Flight path of SO/PHI (blue lines) relative to the Sun (black cross) and SDO/HMI (red cross) over the observation periods of P1, P2, and P3 as seen from above the north pole of the Sun. From this perspective, SO/PHI is moving anti-clockwise about the Sun towards the Sun-SDO/HMI line (black line) and the angle subtended by SO/PHI and SDO/HMI from the Sun represents the separation in heliographic longitude. The solid grey lines mark the annotated angular distances and the dashed grey lines the annotated distances from the Sun. The separation in heliographic latitude, hidden in this perspective, is relatively minute, ranging from $3.7^{\circ}$ to $7.6^{\circ}$. Simultaneous SO/PHI and SDO/HMI magnetograms from 05.09.2021 and 09.09.2021 (blue plus symbols) are depicted in Fig. \ref{armap}.}
\label{flightpath}
\end{figure}

The SO/PHI instrument images the full solar disc with one of its two telescopes, the Full Disc Telescope (FDT). The FDT records polarised full-disc images in bandpasses centred on six wavelength positions, five within the magnetically-sensitive Fe I 6173 \AA{} line, and one in the nearby continuum, on a $2048\times2048$ pixel CMOS sensor. The pixel scale is 3.57 arcsec. From the filtergram observations, various data products are generated, including vector and {line-of-sight} magnetograms, images in the continuum near the Fe I 6173 \AA{} line, and Dopplergrams \citep{kinga20}.

Over the periods of 01.09.2021 to 10.09.2021, 26.09.2021 to 04.10.2021, and 08.10.2021 to 12.10.2021, denoted P1, P2, and P3, the FDT returned an image pair (i.e. simultaneous full-disc {line-of-sight} magnetogram and continuum intensity image) every two hours, with interruptions at certain times. {The Sun-Solar Orbiter distance ranged from about 0.6 au to 0.7 au (Fig. \ref{flightpath}), such that each image pixel corresponded to an area of about $1600\ {\rm km}\times1600\ {\rm km}$ to $1800\ {\rm km}\times1800\ {\rm km}$ on the solar disc. At this distance, the entire solar disc laid within a $1024\times1024$ pixel window in the middle of the field-of-view (FoV).} For this reason, the instrument returned this window alone, producing $1024\times1024$ pixel images. The image pairs with clear instrumental artefacts, such as missing data, are excluded, leaving 173 image pairs, about seven per day on average. At the time of study, the data reduction remains preliminary.

Similar to the FDT on SO/PHI, SDO/HMI records polarised full-disc images in six filters within the Fe I 6173 \AA{} line, from which similar data products are generated at 45-s and 720-s cadences. The CCD/image size is $4096\times4096$ pixel and the pixel scale is 0.5 arcsec, such that each image pixel corresponds to an area of about $360\ {\rm km}\times360\ {\rm km}$ on the solar disc.

We refer, by the term SO/PHI, to the FDT on this instrument. We isolate the SDO/HMI 720-s {line-of-sight} magnetogram and continuum image (hmi.M\_720s and hmi.Ic\_720s data products) concurrent to each SO/PHI image pair taking into account that the light travel time from the Sun to SO/PHI is shorter than that to SDO/HMI. The average difference in the time of observation between the SDO/HMI and SO/PHI image pairs is about four minutes. To match the SDO/HMI data set to the SO/PHI data set, taking each SDO/HMI image, the solar disc is binned down to the size of the solar disc in the corresponding SO/PHI image. Then, we resampled the SDO/HMI image in such a way so as to account for solar rotation in the time interval between the two images. From P1 to P3, SO/PHI approached the Sun-SDO/HMI line such that the angular distance between SO/PHI and SDO/HMI from the Sun declined steadily from $67.3^{\circ}$ to $12.2^{\circ}$ (see Fig. \ref{flightpath}).

\subsection{SATIRE-S model}
\label{satiresmodel}

\begin{figure}
\centering
\includegraphics{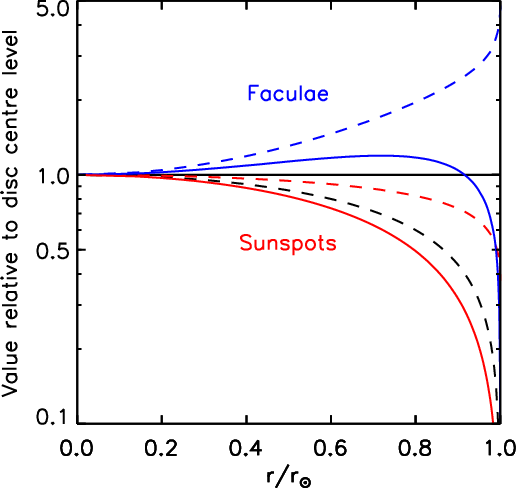}
\caption{{TSI excess produced by a given facular feature in SATIRE-S (blue solid line) and the facular intensity contrast adopted by the model (blue dashed line), from \cite{unruh99}, as a function of distance from disc centre, $\rn$. The red solid and dashed lines represent the same for sunspots. The black dashed line follows the projected area of a given solar surface feature (i.e., foreshortening). Each quantity is normalised to the disc centre value (black solid line).} The model treats sunspot umbra and sunspot penumbra separately, and assumes an umbral to penumbral area ratio of 1:4. For each quantity depicted, the sunspot profile is given by the mean of the umbral and penumbral profiles, weighted by this ratio.}
\label{clv}
\end{figure}

The SATIRE-S model describes the variation in solar irradiance due to sunspot darkening and facular and network brightening. The model does not distinguish between faculae and network, which are collectively termed 'faculae'. The model has two components. The first component is the solar disc coverage by sunspot umbra, sunspot penumbra, and faculae at a given time, derived by identifying these features in the image pair {(line-of-sight magnetogram and continuum intensity image)} recorded then. The solar disc outside of sunspots and faculae is classified as internetwork. The second component is the intensity spectra of sunspot umbra, sunspot penumbra, faculae, and the internetwork at various distances from disc centre, calculated from models of their atmospheric structure with a radiative transfer code \citep{unruh99}. Integrating each intensity spectrum over the full wavelength range, 115 nm to 170,000 nm, yields the corresponding bolometric intensity.

{Hereinafter, we will refer to the difference between facular and internetwork bolometric intensity, depicted in Fig. \ref{clv} (blue dashed line), as the facular intensity contrast. Likewise, we will refer to the difference between sunspot and internetwork bolometric intensity (red dashed line) as the sunspot intensity contrast. Going from disc centre to limb, the facular intensity contrast increases steadily from the disc centre level (black solid line) while the sunspot intensity contrast decreases.} To recover TSI, the model assigns to each point on the solar disc the appropriate bolometric intensity depending on whether it is inside the umbra or penumbra of a sunspot, faculae, or the internetwork and the distance from disc centre. The summation of the result over the solar disc yields TSI. 

{Sunspots are distinguishable in continuum images by the negative intensity contrast and faculae are distinguishable in magnetograms from noise by the elevated magnetogram signals. Let $\blos$ denote the {line-of-sight} magnetogram signal, which indicates the line-of-sight component of the mean magnetic flux density over each resolution element. For a given image pair, the points on the solar disc where the intensity contrast is below a certain threshold are classified as sunspots and the points where $\blosabs$ is above a certain threshold and not already counted as sunspots are taken to correspond to faculae.}

The magnetic concentrations that comprise faculae and network remain largely unresolved at the spatial resolution of available full-disc magnetograms, including SO/PHI and SDO/HMI. In SATIRE-S, the filling factor of facular pixels, that is, the effective proportion of each resolution element occupied by faculae, is estimated by means of an empirical relationship. {Let $\mu$ denote the cosine of the heliocentric angle.} Taking into consideration that magnetic flux tubes tend towards a surface normal orientation due to magnetic buoyancy, the quantity $\blosmu$ approximates the mean magnetic flux density over each resolution element. The faculae filling factor is taken to scale linearly with $\blosmu$ from zero at 0 G to unity at what is denoted $\blosmusat$, above which it saturates. {The $\blosmusat$ parameter modulates the facular area and, thus, the amplitude of the facular brightening as well.}

The TSI excess or deficit produced by a given facular or sunspot feature depends on the projected area and intensity contrast. The projected area changes with its time evolution and its passage across the solar disc with solar rotation (i.e. foreshortening) and the intensity contrast with the latter due to the centre-to-limb variation (CLV) of this quantity. The SATIRE-S model, by establishing the disc-coverage by faculae and sunspots and taking the CLV of facular, sunspot, and internetwork intensity into account {(Fig. \ref{clv})}, captures these processes.

\subsection{TSI reconstruction}
\label{tsireconstruction}

\begin{figure*}
\centering
\includegraphics{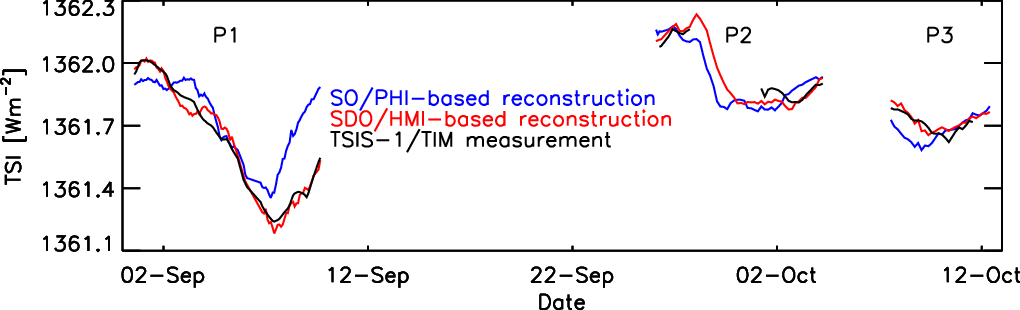}
\caption{Total solar irradiance (TSI) over P1, P2 and P3. The blue and red lines correspond to the SATIRE-S model reconstruction of TSI at 2h cadence based on the SO/PHI and SDO/HMI data sets, respectively. The black line traces the concurrent measurements from {TSIS-1/TIM}: the 6h cadence record is interpolated to the {times} of the model reconstruction.}
\label{tsi}
\end{figure*}

{We apply the SATIRE-S model to the SDO/HMI and SO/PHI data sets to reconstruct TSI variability as apparent from both perspectives. In the model, the intensity contrast threshold determines the disc coverage by sunspots while the $\blosabs$ threshold and $\blosmusat$ determine the disc coverage by faculae.} While SO/PHI and SDO/HMI survey the same spectral line, they do differ in instrument design and data reduction, most notably, the significant difference in spatial resolution (Sect. \ref{dataselectionreduction}). The result is systematic differences in the apparent sunspot intensity contrast, magnetogram noise, and magnetogram signal scale between the two data sets. {The intensity contrast and $\blosabs$ thresholds and $\blosmusat$ have to be adapted to each data set to account for these differences. In earlier implementations of the SATIRE-S model \citep{krivova03,yeo14a}, the intensity contrast threshold is set at the level that brings the resultant sunspot area into agreement with independent sunspot area measurements, and the $\blosabs$ threshold is set at the 3$\sigma$ magnetogram noise level. The $\blosmusat$ parameter is set at the level that matches the amplitude of TSI variability in the model output to that in TSI measurements, making it a free parameter of the model. However, available sunspot area and TSI measurements are all recorded from the Earth's perspective, meaning we cannot apply the above procedure as is to the SO/PHI data set. We modified the procedure of \cite{krivova03} and \cite{yeo14a} in such a way so we can set the various parameters at the appropriate level for each data set, while circumventing the issue that there are no sunspot area and TSI measurements from PHI's perspective. The objective is to ensure that the disc coverage by faculae and sunspots is derived from the SDO/HMI and SO/PHI data sets consistently; in turn, this ensures that the model output based on the two are consistent.} In the following analysis, unless specified, we make use of all available data, namely, P1, P2, and P3, in order to maximise the statistics.

The continuum images were normalised by their limb-darkening profile, determined following \cite{neckel94}. Taking each SDO/HMI (SO/PHI) continuum image, we classified the points on the solar disc where the normalised intensity lies below 0.91 (0.94) as sunspots. For SDO/HMI, the threshold is set at 0.91 by finding the level where the resultant disc-integrated sunspot area complies with the independent record by \cite{mandal20}. The fact that SO/PHI is outside the Sun-Earth line excludes a similar comparison to available sunspot area measurements. Instead, the threshold was set at 0.94 by finding the level where the foreshortening-corrected sunspot area measured by SO/PHI matches that by SDO/HMI over the part of the solar surface imaged by both instruments. The sunspot pixels within each image are ordered by the normalised intensity and the points below (above) the first quintile are taken to correspond to sunspot umbra (penumbra). This sets the umbral to penumbral area ratio at 1:4, in compliance with independent sunspot observations \citep{solanki03}.

Taking each SDO/HMI (SO/PHI) magnetogram, the points on the solar disc where $\blosabs$ is above 45 G (31 G) and not already identified as sunspots are taken to correspond to faculae. Standalone facular pixels are reclassified as internetwork to minimise the wrongful inclusion of magnetogram noise as faculae. We determined the appropriate $\blos$ threshold to separate faculae from magnetogram noise by the following procedure.

For SO/PHI, we determined the magnetogram noise level as a function of position in the FOV following the procedure of \cite{ortiz02} and \cite{yeo13}. This revealed the 1$\sigma$ noise level to vary within the range of 5.0 G and 10.2 G. Conservatively, we set the $\blos$ threshold for SO/PHI at 31 G, three times the upper limit of this range. Compared to the SO/PHI magnetograms, the SDO/HMI magnetograms are less noisy, due mainly to the longer integration time and the fact that we had resampled them so that the size of the solar disc matches that in the former (Sect. \ref{dataselectionreduction}). Consequently, there are weak facular features that are distinguishable from noise in the SDO/HMI magnetograms but obscured by noise in the SO/PHI magnetograms. Therefore, instead of setting the $\blos$ threshold for SDO/HMI at its 3$\sigma$ noise level like we did for SO/PHI, we set it at the level that ensures we are isolating only the facular features that SO/PHI can also distinguish from noise. Confining ourselves to P3, we rotated the solar disc in the SDO/HMI magnetograms to SO/PHI's perspective. {We let $\mu_{\rm HMI}$ and $\mu_{\rm PHI}$ denote, for a given point on the solar surface, the value of $\mu$ in SDO/HMI's and in SO/PHI's perspective, respectively.} We scaled each point on the solar disc in the rotated SDO/HMI magnetogram by $\frac{\mu_{\rm PHI}}{\mu_{\rm HMI}}$. We refer to the result (an estimate of what the SDO/HMI instrument would have recorded if it was in SO/PHI's position) as the projected SDO/HMI magnetograms. This step is more reliable when the two perspectives are closer, which is why we confined this analysis to P3. We set the $\blos$ threshold for SDO/HMI at 45 G by finding the value where the number of facular pixels in the projected SDO/HMI magnetograms matches that in the SO/PHI magnetograms over the part of the solar surface imaged by both instruments.

The free parameter in the model, $\blosmusat$ is set at 94 G for the SO/PHI-based reconstruction and at 146 G for the SDO/HMI-based reconstruction. Recall, $\blosmusat$ modulates the amplitude of facular brightening (Sect. \ref{satiresmodel}). For the SDO/HMI-based reconstruction, $\blosmusat$ is constrained by matching the amplitude of TSI variability in the model output to that in observed TSI. For this purpose, we make use of the TSI record from the Total Irradiance Monitor (TIM) on the Total and Spectral Solar Irradiance Sensor ({TSIS-1}) suite onboard the International Space Station\footnote{Version 3, available at lasp.colorado.edu/home/tsis/data/}. We emphasise that by this step, the SDO/HMI-based reconstruction (red, Fig. \ref{tsi}) merely adopts the amplitude of TSI variability in the {TSIS-1/TIM} record (black). No adjustment of the free parameter can change the variation in TSI with time in the model output to match that in the {TSIS-1/TIM} record. As illustrated, the SDO/HMI-based reconstruction closely reproduces the {TSIS-1/TIM} record: the Pearson's correlation coefficient, $R^2$ is 0.981, and the root-mean-square difference, RMSD, is $0.0361\ {\rm Wm^{-2}}$. The fact that SO/PHI is outside the Sun-Earth line excludes a similar comparison to available TSI measurements. To circumvent this issue, again we compared the SO/PHI magnetograms from P3 to the result of projecting the corresponding SDO/HMI magnetograms to SO/PHI's perspective. For the SO/PHI-based reconstruction, $\blosmusat$ is constrained by matching the disc-integrated faculae filling factor in the SO/PHI magnetograms to that in the projected SDO/HMI magnetograms over the part of the solar surface imaged by both instruments. {The SO/PHI-based reconstruction is shown in Fig. \ref{tsi} (blue) along with the SDO/HMI-based reconstruction (red) and the TSIS-1/TIM record (black). Of course, due to the difference in perspective, the SO/PHI time series is not expected to match the other two time series.}

\section{Results}
\label{section3}

\begin{figure*}
\centering
\includegraphics{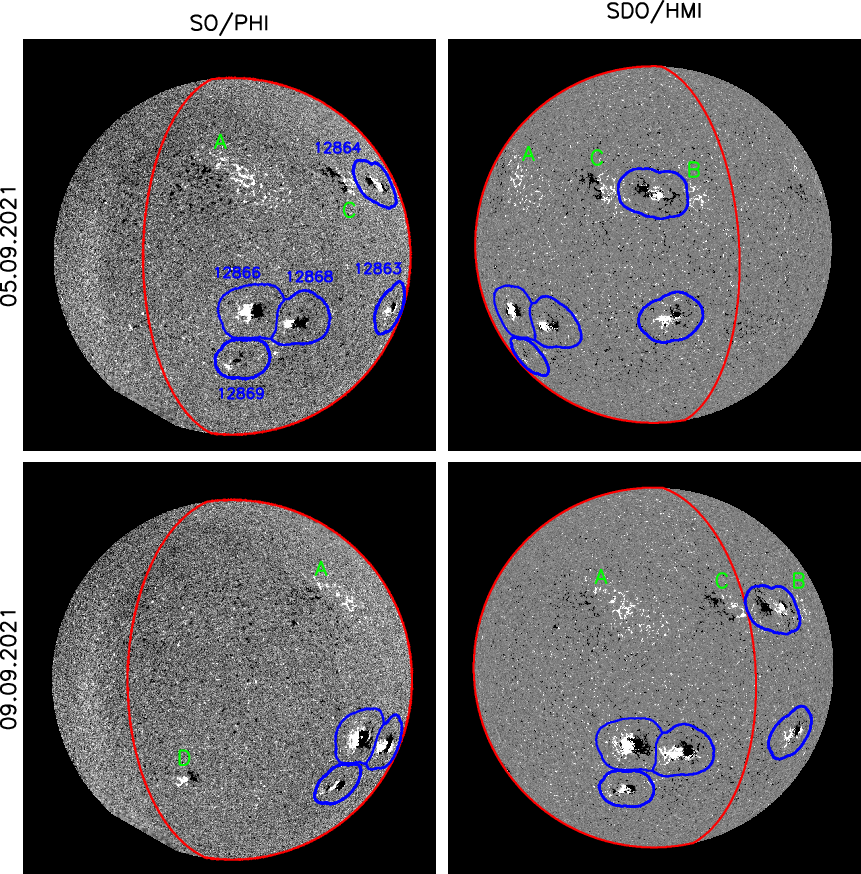}
\caption{Simultaneous SO/PHI (left) and SDO/HMI {line-of-sight} magnetogram (right) recorded during P1 on 05.09.2021 (top), and 09.09.2021 (bottom). The position of SO/PHI relative to the Sun-SDO/HMI line at these times is indicated in Fig. \ref{flightpath} (blue plus symbols). The part of the solar surface within view of both instruments is encircled in red. The five active regions (ARs) that emerged within P1 and were imaged by both instruments are encircled in blue. The NOAA AR number is indicated in the top-left panel. The magnetic features marked "A" and "B" correspond to remnants of ARs which dissipated before P1, and "C" and "D" to ARs which do not meet the aforementioned criteria, discussed in Appendix \ref{appendixar}. The greyscale is saturated for the SO/PHI and SDO/HMI magnetograms at $\pm31\ {\rm G}$ and $\pm45\ {\rm G}$, respectively. This corresponds to the magnetogram signal threshold applied to separate faculae from magnetogram noise (Sect. \ref{tsireconstruction}).}
\label{armap}
\end{figure*}

\begin{figure*}
\centering
\includegraphics{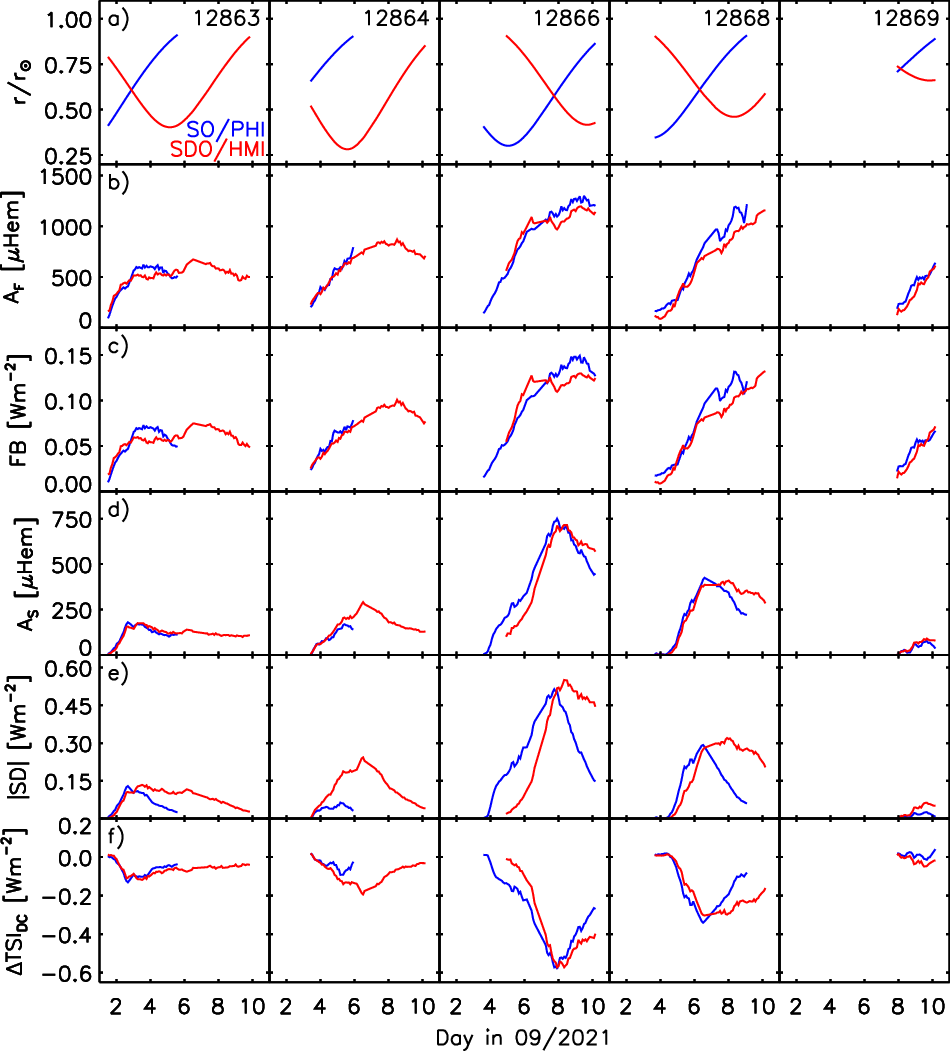}
\caption{Properties of the five active regions that emerged during P1 and imaged by both SO/PHI and SDO/HMI (outlined by blue contours in Fig. \ref{armap}), as a function of time: a) the distance from disc centre, $\rn$, b) foreshortening-corrected facular area, $\af$ in millionths of the solar hemisphere, $\mu{\rm Hem}$, c) the change in TSI due to faculae, $\fb$. {Rows d) and e) represent the same as rows b) and c), except for sunspots. In row f), we chart the total change in TSI (i.e. both faculae brightening and sunspot darkening) produced by each active region to an observer that follows it such that it is always at disc centre, ${\rm \Delta{}TSI_{DC}}$.} The blue (red) solid lines follow the values from the SO/PHI (SDO/HMI) data set and the corresponding TSI reconstruction. See Sect. \ref{section3} for details.}
\label{artsi}
\end{figure*}

{As discussed in the previous section, we constrained the SO/PHI-based model by optimising the agreement between the disc coverage by faculae and sunspots derived from the SO/PHI and SDO/HMI data sets (Sect. \ref{tsireconstruction}). Of course, even with such a step, there will be residual discrepancies between the disc coverage by faculae and sunspots derived from the two data sets. In this section, we examine the extent to which faculae and sunspots are mapped consistently between the two data sets, and the uncertainty in the model output TSI produced by the inevitable uncertainty in the disc coverage by faculae and sunspots.}

{To this end, we identify the active regions (ARs) that were imaged by both instruments and examine the area of the enclosed faculae and sunspots and their contribution to the TSI variability reproduced by the model. We confine this analysis to P1, where the two perspectives are furthest apart. In Sect. \ref{tsireconstruction}, we made use of the data from P3, where the two perspectives are closest to one another, to constrain the $\blos$ threshold and $\blosmusat$, setting them at the levels that brought the disc coverage by faculae in the two data sets into agreement over this period. There is a need to verify that faculae continue to be mapped consistently between the two data sets outside of this period, especially when the two perspectives are far apart.}

Taking the SO/PHI and SDO/HMI data sets, we identified and established the boundary of ARs that had emerged during P1 and were imaged by both instruments (Appendix \ref{appendixar}), isolating five such ARs (blue contours, Fig. \ref{armap}). {We examine each AR over the period where it contains sunspots and lies entirely within the solar disc from a given perspective.} We let $\rn$ denote the distance of the magnetic centre-of-mass of a given AR from disc centre, normalised to the solar radius. {The trajectory of the various ARs across the solar disc as seen from both perspectives, in terms of $\rn$, is given in Fig. \ref{artsi}a. Compared to P1, periods P2 and P3 are shorter and solar activity was weaker such that there are only two relatively small and weak emerging ARs imaged by both instruments and even then only for a limited period of time. This is another reason the current analysis is confined to P1.}

{From the disc-coverage by faculae and sunspots determined from the SO/PHI and SDO/HMI data sets, we derive the foreshortening-corrected facular ($\af$, Fig. \ref{artsi}b) and sunspot area within each AR ($\as$, Fig. \ref{artsi}d). Comparing the $\af$ values from the SO/PHI (blue) and SDO/HMI data sets (red), the $R^2$ is 0.956 and the RMSD is 88.9 $\mu$Hem. In the case of $\as$, the $R^2$ is 0.939 and the RMSD is 52.3 $\mu$Hem. The values from the two data sets are fairly similar, verifying the steps in Sect. \ref{tsireconstruction} to ensure we map faculae and sunspots consistently between the two data sets.}

For each AR and from each perspective, we compute the contribution to TSI variability by the enclosed faculae (facular brightening, $\fb$, Fig. \ref{artsi}c) and by the enclosed sunspots (sunspot darkening, $\sd$, Fig. \ref{artsi}e). To derive the $\fb$ of a given AR, we repeat the TSI reconstruction, treating the faculae within this AR as internetwork and subtracting the result from the original reconstruction, likewise for $\sd$. {The discrepancies between the $\fb$ and $\sd$ values from the SO/PHI-based (blue) and SDO/HMI-based reconstructions (red) is due to the uncertainty in the disc coverage by faculae and sunspots, just discussed, and rotational modulation. The trajectory of the AR across the solar disc (Fig. \ref{artsi}a), and therefore the effect of rotational modulation on $\fb$ and $\sd$, differs between the two perspectives.}

{Next, we examine the discrepancy between the SO/PHI-based and SDO/HMI-based reconstructions due to the uncertainty in the disc coverage by faculae and sunspots. First, we have to account for the differences between the two due to rotational modulation. Let $\fbdc$ and $\sddc$ denote, for a given AR, the $\fb$ and $\sd$ to an observer that follows it such that it is always at disc centre, namely, no rotational modulation.} In Fig. \ref{clv}, we chart the CLV of facular intensity contrast in SATIRE-S (blue dashed line), based on \cite{unruh99}, and of the projected area of a given solar surface feature (i.e. foreshortening, black dashed line), normalised to the disc centre level (black solid line). The product of the two normalised CLV profiles yields the normalised CLV of the $\fb$ produced by a given facular feature in the model (blue solid line). In the same manner, we derive the normalised CLV of the $\sd$ produced by a given sunspot (red solid line). {For faculae, the two effects roughly cancel out, such that the CLV of $\fb$ is weak, explaining why the $\fb$ produced by the various ARs is similar from both perspectives, as illustrated in Fig. \ref{artsi}c. In contrast, for sunspots, the CLV of the two effects are quantitatively similar such that $\sd$, in absolute terms, exhibits a monotonic and increasingly steep decline towards the limb. This, together with the uncertainty in $\as$, resulted in the discrepancy, between the two perspectives, of the $\sd$ produced by the various ARs (Fig. \ref{artsi}e). We use these results to estimate $\fbdc$ and $\sddc$ from $\fb$ and $\sd$. Finally, for each AR, we took the sum of $\fbdc$ and $\sddc$, yielding the TSI variability produced by the AR to an observer that follows it such that it is always at disc centre, ${\rm \Delta{}TSI_{DC}}$ (Fig. \ref{artsi}f). Comparing the ${\rm \Delta{}TSI_{DC}}$ values from the SO/PHI-based (blue) and SDO/HMI-based reconstructions (red), the $R^2$ is 0.864 and the RMSD is $0.0638\ {\rm Wm^{-2}}$.}

{Between the SO/PHI and SDO/HMI data sets, the disc coverages by faculae and sunspots, at least within ARs, agree to about 4\% and 6\%, respectively, and this corresponds to an agreement of about 86\% between the SO/PHI-based and SDO/HMI-based reconstructions of TSI variability.}

\section{Conclusion}
\label{conclusion}

{In this study, we presented a method to combine the unprecedented observations of the photospheric magnetic field and continuum intensity that SO/PHI is recording from outside the Sun-Earth line with solar observations recorded from the Earth's perspective to examine solar irradiance variability from both perspectives simultaneously.} Taking SO/PHI observations from the cruise phase of the Solar Orbiter mission and concurrent observations from SDO/HMI as input into the SATIRE-S model, we successfully reconstructed total solar irradiance (TSI) variability as apparent from both perspectives (Sects. \ref{section2} and \ref{section3}).

In later stages of the Solar Orbiter mission, the orbital plane will tilt in such a way so as to bring the spacecraft away from the ecliptic to heliographic latitudes of up to $33^{\circ}$. The current study sets the template for the reconstruction of solar irradiance variability as seen from outside the ecliptic from the data SO/PHI is expected to collect from such positions. {In the absence of out-of-ecliptic observations of the Sun, }such a reconstruction is missing from existing attempts to factor inclination into how the brightness variations of the Sun compares to that of other cool stars, which come in all inclinations \citep{schatten93,knaack01,vieira12,shapiro14,nemec20a,nemec20b}.

This study is based on a relative small body of SO/PHI and SDO/HMI observations{ and} the reduction of the SO/PHI data is preliminary. The results obtained here will have to be confirmed, in future efforts, with a larger set of SO/PHI and SDO/HMI observations and the finalised reduction of the former. Future efforts will also aim to reconstruct solar brightness variability in the passbands employed in stellar surveys which monitor the photometry of their program stars, for example, CoRot \citep{auvergne09}, Kepler \citep{borucki10}, TESS \citep{ricker15}, and PLATO \citep{rauer14}.

\begin{acknowledgements}
{The authors would like to thank the referee, Greg Kopp for his invaluable comments on the manuscript.} KLY and NAK received funding from BMBF (project 01LG1909C), and KLY and SKS from ERC through the Horizon 2020 research and innovation program of the EU (grant 695075). The Solar Orbiter mission is an international cooperation between ESA and NASA operated by ESA. We are grateful to the ESA SOC and MOC teams for their support. The German contribution to SO/PHI is funded by BMWK through DLR (grants 50 OT 1001/1201/1901 and 50 OT 0801/1003/1203/1703) and by MPG through its central funds. The Spanish contribution is funded by AEI/MCIN/10.13039/501100011033 (RTI2018-096886-C5, PID2021-125325OB-C5, PCI2022-135009-2), 'ERDF A way of making Europe´, the Severo Ochoa Center of Excellence accreditation awarded to IAA-CSIC (SEV2017-0709, CEX2021-001131-S) and a Ram\'{o}n y Cajal fellowship awarded to DOS. The French contribution is funded by CNES. The HMI data are courtesy of NASA/SDO and the HMI science team. 
\end{acknowledgements}

\bibliographystyle{aa}
\bibliography{references}

\begin{thebibliography}{37}
\expandafter\ifx\csname natexlab\endcsname\relax\def\natexlab#1{#1}\fi

\bibitem[{{Albert} {et~al.}(2020){Albert}, {Hirzberger}, {Kolleck}, {Jorge},
  {Busse}, {Rodr{\'\i}guez}, {Carrascosa}, {Fiethe}, {Gandorfer}, {Germerott},
  {Guan}, {Guerrero}, {Gutierrez-Marques}, {Exp{\'o}sito}, {Lange}, {Michalik},
  {Su{\'a}rez}, {Schou}, {Solanki}, {del Toro Iniesta}, \& {Woch}}]{kinga20}
{Albert}, K., {Hirzberger}, J., {Kolleck}, M., {et~al.} 2020, J. Astron.
  Telesc. Instrum. Syst., 6, 048004

\bibitem[{{Auvergne} {et~al.}(2009){Auvergne}, {Bodin}, {Boisnard}, {Buey},
  {Chaintreuil}, {Epstein}, {Jouret}, {Lam-Trong}, {Levacher}, {Magnan},
  {Perez}, {Plasson}, {Plesseria}, {Peter}, {Steller}, {Tiph{\`e}ne}, {Baglin},
  {Agogu{\'e}}, {Appourchaux}, {Barbet}, {Beaufort}, {Bellenger}, {Berlin},
  {Bernardi}, {Blouin}, {Boumier}, {Bonneau}, {Briet}, {Butler}, {Cautain},
  {Chiavassa}, {Costes}, {Cuvilho}, {Cunha-Parro}, {de Oliveira Fialho},
  {Decaudin}, {Defise}, {Djalal}, {Docclo}, {Drummond}, {Dupuis}, {Exil},
  {Faur{\'e}}, {Gaboriaud}, {Gamet}, {Gavalda}, {Grolleau}, {Gueguen},
  {Guivarc'h}, {Guterman}, {Hasiba}, {Huntzinger}, {Hustaix}, {Imbert},
  {Jeanville}, {Johlander}, {Jorda}, {Journoud}, {Karioty}, {Kerjean},
  {Lafond}, {Lapeyrere}, {Landiech}, {Larqu{\'e}}, {Laudet}, {Le Merrer},
  {Leporati}, {Leruyet}, {Levieuge}, {Llebaria}, {Martin}, {Mazy}, {Mesnager},
  {Michel}, {Moalic}, {Monjoin}, {Naudet}, {Neukirchner}, {Nguyen-Kim},
  {Ollivier}, {Orcesi}, {Ottacher}, {Oulali}, {Parisot}, {Perruchot},
  {Piacentino}, {Pinheiro da Silva}, {Platzer}, {Pontet}, {Pradines},
  {Quentin}, {Rohbeck}, {Rolland}, {Rollenhagen}, {Romagnan}, {Russ}, {Samadi},
  {Schmidt}, {Schwartz}, {Sebbag}, {Smit}, {Sunter}, {Tello}, {Toulouse},
  {Ulmer}, {Vandermarcq}, {Vergnault}, {Wallner}, {Waultier}, \&
  {Zanatta}}]{auvergne09}
{Auvergne}, M., {Bodin}, P., {Boisnard}, L., {et~al.} 2009, \aap, 506, 411

\bibitem[{{Borucki} {et~al.}(2010){Borucki}, {Koch}, {Basri}, {Batalha},
  {Brown}, {Caldwell}, {Caldwell}, {Christensen-Dalsgaard}, {Cochran},
  {DeVore}, {Dunham}, {Dupree}, {Gautier}, {Geary}, {Gilliland}, {Gould},
  {Howell}, {Jenkins}, {Kondo}, {Latham}, {Marcy}, {Meibom}, {Kjeldsen},
  {Lissauer}, {Monet}, {Morrison}, {Sasselov}, {Tarter}, {Boss}, {Brownlee},
  {Owen}, {Buzasi}, {Charbonneau}, {Doyle}, {Fortney}, {Ford}, {Holman},
  {Seager}, {Steffen}, {Welsh}, {Rowe}, {Anderson}, {Buchhave}, {Ciardi},
  {Walkowicz}, {Sherry}, {Horch}, {Isaacson}, {Everett}, {Fischer}, {Torres},
  {Johnson}, {Endl}, {MacQueen}, {Bryson}, {Dotson}, {Haas}, {Kolodziejczak},
  {Van Cleve}, {Chandrasekaran}, {Twicken}, {Quintana}, {Clarke}, {Allen},
  {Li}, {Wu}, {Tenenbaum}, {Verner}, {Bruhweiler}, {Barnes}, \&
  {Prsa}}]{borucki10}
{Borucki}, W.~J., {Koch}, D., {Basri}, G., {et~al.} 2010, Science, 327, 977

\bibitem[{{Chatzistergos} {et~al.}(2022){Chatzistergos}, {Krivova}, \&
  {Ermolli}}]{chatzistergos22}
{Chatzistergos}, T., {Krivova}, N.~A., \& {Ermolli}, I. 2022, Front. Astron.
  Space Sci., 9, 1038949

\bibitem[{{Domingo} {et~al.}(2009){Domingo}, {Ermolli}, {Fox}, {Fr{\"o}hlich},
  {Haberreiter}, {Krivova}, {Kopp}, {Schmutz}, {Solanki}, {Spruit}, {Unruh}, \&
  {V{\"o}gler}}]{domingo09}
{Domingo}, V., {Ermolli}, I., {Fox}, P., {et~al.} 2009, \ssr, 145, 337

\bibitem[{{Ermolli} {et~al.}(2013){Ermolli}, {Matthes}, {Dudok de Wit},
  {Krivova}, {Tourpali}, {Weber}, {Unruh}, {Gray}, {Langematz}, {Pilewskie},
  {Rozanov}, {Schmutz}, {Shapiro}, {Solanki}, \& {Woods}}]{ermolli13}
{Ermolli}, I., {Matthes}, K., {Dudok de Wit}, T., {et~al.} 2013, Atmos. Chem.
  Phys., 13, 3945

\bibitem[{{Fligge} {et~al.}(2000){Fligge}, {Solanki}, \& {Unruh}}]{fligge00}
{Fligge}, M., {Solanki}, S.~K., \& {Unruh}, Y.~C. 2000, \aap, 353, 380

\bibitem[{{Foukal}(1993)}]{foukal93}
{Foukal}, P. 1993, \solphys, 148, 219

\bibitem[{{Gray} {et~al.}(2010){Gray}, {Beer}, {Geller}, {Haigh}, {Lockwood},
  {Matthes}, {Cubasch}, {Fleitmann}, {Harrison}, {Hood}, {Luterbacher},
  {Meehl}, {Shindell}, {van Geel}, \& {White}}]{gray10}
{Gray}, L.~J., {Beer}, J., {Geller}, M., {et~al.} 2010, Rev. Geophys., 48,
  RG4001

\bibitem[{{Jakosky} {et~al.}(2015){Jakosky}, {Lin}, {Grebowsky}, {Luhmann},
  {Mitchell}, {Beutelschies}, {Priser}, {Acuna}, {Andersson}, {Baird}, {Baker},
  {Bartlett}, {Benna}, {Bougher}, {Brain}, {Carson}, {Cauffman}, {Chamberlin},
  {Chaufray}, {Cheatom}, {Clarke}, {Connerney}, {Cravens}, {Curtis}, {Delory},
  {Demcak}, {DeWolfe}, {Eparvier}, {Ergun}, {Eriksson}, {Espley}, {Fang},
  {Folta}, {Fox}, {Gomez-Rosa}, {Habenicht}, {Halekas}, {Holsclaw}, {Houghton},
  {Howard}, {Jarosz}, {Jedrich}, {Johnson}, {Kasprzak}, {Kelley}, {King},
  {Lankton}, {Larson}, {Leblanc}, {Lefevre}, {Lillis}, {Mahaffy}, {Mazelle},
  {McClintock}, {McFadden}, {Mitchell}, {Montmessin}, {Morrissey}, {Peterson},
  {Possel}, {Sauvaud}, {Schneider}, {Sidney}, {Sparacino}, {Stewart}, {Tolson},
  {Toublanc}, {Waters}, {Woods}, {Yelle}, \& {Zurek}}]{jakosky15}
{Jakosky}, B.~M., {Lin}, R.~P., {Grebowsky}, J.~M., {et~al.} 2015, \ssr, 195, 3

\bibitem[{{Knaack} {et~al.}(2001){Knaack}, {Fligge}, {Solanki}, \&
  {Unruh}}]{knaack01}
{Knaack}, R., {Fligge}, M., {Solanki}, S.~K., \& {Unruh}, Y.~C. 2001, \aap,
  376, 1080

\bibitem[{{Kopp}(2014)}]{kopp14}
{Kopp}, G. 2014, J. Space Weather Space Clim., 4, A14

\bibitem[{{Krivova} {et~al.}(2003){Krivova}, {Solanki}, {Fligge}, \&
  {Unruh}}]{krivova03}
{Krivova}, N.~A., {Solanki}, S.~K., {Fligge}, M., \& {Unruh}, Y.~C. 2003, \aap,
  399, L1

\bibitem[{{Mandal} {et~al.}(2020){Mandal}, {Krivova}, {Solanki}, {Sinha}, \&
  {Banerjee}}]{mandal20}
{Mandal}, S., {Krivova}, N.~A., {Solanki}, S.~K., {Sinha}, N., \& {Banerjee},
  D. 2020, \aap, 640, A78

\bibitem[{{M{\"u}ller} {et~al.}(2020){M{\"u}ller}, {St. Cyr}, {Zouganelis},
  {Gilbert}, {Marsden}, {Nieves-Chinchilla}, {Antonucci}, {Auch{\`e}re},
  {Berghmans}, {Horbury}, {Howard}, {Krucker}, {Maksimovic}, {Owen}, {Rochus},
  {Rodriguez-Pacheco}, {Romoli}, {Solanki}, {Bruno}, {Carlsson}, {Fludra},
  {Harra}, {Hassler}, {Livi}, {Louarn}, {Peter}, {Sch{\"u}hle}, {Teriaca}, {del
  Toro Iniesta}, {Wimmer-Schweingruber}, {Marsch}, {Velli}, {De Groof},
  {Walsh}, \& {Williams}}]{muller20}
{M{\"u}ller}, D., {St. Cyr}, O.~C., {Zouganelis}, I., {et~al.} 2020, \aap, 642,
  A1

\bibitem[{{Neckel} \& {Labs}(1994)}]{neckel94}
{Neckel}, H. \& {Labs}, D. 1994, \solphys, 153, 91

\bibitem[{{N{\`e}mec} {et~al.}(2020{\natexlab{a}}){N{\`e}mec}, {I{\c{s}}{\i}k},
  {Shapiro}, {Solanki}, {Krivova}, \& {Unruh}}]{nemec20b}
{N{\`e}mec}, N.~E., {I{\c{s}}{\i}k}, E., {Shapiro}, A.~I., {et~al.}
  2020{\natexlab{a}}, \aap, 638, A56

\bibitem[{{N{\`e}mec} {et~al.}(2020{\natexlab{b}}){N{\`e}mec}, {Shapiro},
  {Krivova}, {Solanki}, {Tagirov}, {Cameron}, \& {Dreizler}}]{nemec20a}
{N{\`e}mec}, N.~E., {Shapiro}, A.~I., {Krivova}, N.~A., {et~al.}
  2020{\natexlab{b}}, \aap, 636, A43

\bibitem[{{Ortiz} {et~al.}(2002){Ortiz}, {Solanki}, {Domingo}, {Fligge}, \&
  {Sanahuja}}]{ortiz02}
{Ortiz}, A., {Solanki}, S.~K., {Domingo}, V., {Fligge}, M., \& {Sanahuja}, B.
  2002, \aap, 388, 1036

\bibitem[{{Radick} {et~al.}(2018){Radick}, {Lockwood}, {Henry}, {Hall}, \&
  {Pevtsov}}]{radick18}
{Radick}, R.~R., {Lockwood}, G.~W., {Henry}, G.~W., {Hall}, J.~C., \&
  {Pevtsov}, A.~A. 2018, \apj, 855, 75

\bibitem[{{Rauer} {et~al.}(2014){Rauer}, {Catala}, {Aerts}, {Appourchaux},
  {Benz}, {Brandeker}, {Christensen-Dalsgaard}, {Deleuil}, {Gizon}, {Goupil},
  {G{\"u}del}, {Janot-Pacheco}, {Mas-Hesse}, {Pagano}, {Piotto}, {Pollacco},
  {Santos}, {Smith}, {Su{\'a}rez}, {Szab{\'o}}, {Udry}, {Adibekyan}, {Alibert},
  {Almenara}, {Amaro-Seoane}, {Eiff}, {Asplund}, {Antonello}, {Barnes},
  {Baudin}, {Belkacem}, {Bergemann}, {Bihain}, {Birch}, {Bonfils}, {Boisse},
  {Bonomo}, {Borsa}, {Brand{\~a}o}, {Brocato}, {Brun}, {Burleigh}, {Burston},
  {Cabrera}, {Cassisi}, {Chaplin}, {Charpinet}, {Chiappini}, {Church},
  {Csizmadia}, {Cunha}, {Damasso}, {Davies}, {Deeg}, {D{\'\i}az}, {Dreizler},
  {Dreyer}, {Eggenberger}, {Ehrenreich}, {Eigm{\"u}ller}, {Erikson}, {Farmer},
  {Feltzing}, {de Oliveira Fialho}, {Figueira}, {Forveille}, {Fridlund},
  {Garc{\'\i}a}, {Giommi}, {Giuffrida}, {Godolt}, {Gomes da Silva}, {Granzer},
  {Grenfell}, {Grotsch-Noels}, {G{\"u}nther}, {Haswell}, {Hatzes},
  {H{\'e}brard}, {Hekker}, {Helled}, {Heng}, {Jenkins}, {Johansen},
  {Khodachenko}, {Kislyakova}, {Kley}, {Kolb}, {Krivova}, {Kupka}, {Lammer},
  {Lanza}, {Lebreton}, {Magrin}, {Marcos-Arenal}, {Marrese}, {Marques},
  {Martins}, {Mathis}, {Mathur}, {Messina}, {Miglio}, {Montalban}, {Montalto},
  {Monteiro}, {Moradi}, {Moravveji}, {Mordasini}, {Morel}, {Mortier},
  {Nascimbeni}, {Nelson}, {Nielsen}, {Noack}, {Norton}, {Ofir}, {Oshagh},
  {Ouazzani}, {P{\'a}pics}, {Parro}, {Petit}, {Plez}, {Poretti}, {Quirrenbach},
  {Ragazzoni}, {Raimondo}, {Rainer}, {Reese}, {Redmer}, {Reffert},
  {Rojas-Ayala}, {Roxburgh}, {Salmon}, {Santerne}, {Schneider}, {Schou},
  {Schuh}, {Schunker}, {Silva-Valio}, {Silvotti}, {Skillen}, {Snellen}, {Sohl},
  {Sousa}, {Sozzetti}, {Stello}, {Strassmeier}, {{\v{S}}vanda}, {Szab{\'o}},
  {Tkachenko}, {Valencia}, {Van Grootel}, {Vauclair}, {Ventura}, {Wagner},
  {Walton}, {Weingrill}, {Werner}, {Wheatley}, \& {Zwintz}}]{rauer14}
{Rauer}, H., {Catala}, C., {Aerts}, C., {et~al.} 2014, Exp. Astron., 38, 249

\bibitem[{{Ricker} {et~al.}(2015){Ricker}, {Winn}, {Vanderspek}, {Latham},
  {Bakos}, {Bean}, {Berta-Thompson}, {Brown}, {Buchhave}, {Butler}, {Butler},
  {Chaplin}, {Charbonneau}, {Christensen-Dalsgaard}, {Clampin}, {Deming},
  {Doty}, {De Lee}, {Dressing}, {Dunham}, {Endl}, {Fressin}, {Ge}, {Henning},
  {Holman}, {Howard}, {Ida}, {Jenkins}, {Jernigan}, {Johnson}, {Kaltenegger},
  {Kawai}, {Kjeldsen}, {Laughlin}, {Levine}, {Lin}, {Lissauer}, {MacQueen},
  {Marcy}, {McCullough}, {Morton}, {Narita}, {Paegert}, {Palle}, {Pepe},
  {Pepper}, {Quirrenbach}, {Rinehart}, {Sasselov}, {Sato}, {Seager},
  {Sozzetti}, {Stassun}, {Sullivan}, {Szentgyorgyi}, {Torres}, {Udry}, \&
  {Villasenor}}]{ricker15}
{Ricker}, G.~R., {Winn}, J.~N., {Vanderspek}, R., {et~al.} 2015, J. Astron.
  Telesc. Instrum. Syst., 1, 014003

\bibitem[{{Schatten}(1993)}]{schatten93}
{Schatten}, K.~H. 1993, \jgr, 98, 18907

\bibitem[{{Scherrer} {et~al.}(2012){Scherrer}, {Schou}, {Bush}, {Kosovichev},
  {Bogart}, {Hoeksema}, {Liu}, {Duvall}, {Zhao}, {Title}, {Schrijver},
  {Tarbell}, \& {Tomczyk}}]{scherrer12}
{Scherrer}, P.~H., {Schou}, J., {Bush}, R.~I., {et~al.} 2012, \solphys, 275,
  207

\bibitem[{{Shapiro} {et~al.}(2017){Shapiro}, {Solanki}, {Krivova}, {Cameron},
  {Yeo}, \& {Schmutz}}]{shapiro17}
{Shapiro}, A.~I., {Solanki}, S.~K., {Krivova}, N.~A., {et~al.} 2017, Nat.
  Astron., 1, 612

\bibitem[{{Shapiro} {et~al.}(2014){Shapiro}, {Solanki}, {Krivova}, {Schmutz},
  {Ball}, {Knaack}, {Rozanov}, \& {Unruh}}]{shapiro14}
{Shapiro}, A.~I., {Solanki}, S.~K., {Krivova}, N.~A., {et~al.} 2014, \aap, 569,
  A38

\bibitem[{{Solanki}(2003)}]{solanki03}
{Solanki}, S.~K. 2003, \aapr, 11, 153

\bibitem[{{Solanki} {et~al.}(2020){Solanki}, {del Toro Iniesta}, {Woch},
  {Gandorfer}, {Hirzberger}, {Alvarez-Herrero}, {Appourchaux}, {Mart{\'\i}nez
  Pillet}, {P{\'e}rez-Grande}, {Sanchis Kilders}, {Schmidt}, {G{\'o}mez Cama},
  {Michalik}, {Deutsch}, {Fernandez-Rico}, {Grauf}, {Gizon}, {Heerlein},
  {Kolleck}, {Lagg}, {Meller}, {M{\"u}ller}, {Sch{\"u}hle}, {Staub}, {Albert},
  {Alvarez Copano}, {Beckmann}, {Bischoff}, {Busse}, {Enge}, {Frahm},
  {Germerott}, {Guerrero}, {L{\"o}ptien}, {Meierdierks}, {Oberdorfer},
  {Papagiannaki}, {Ramanath}, {Schou}, {Werner}, {Yang}, {Zerr}, {Bergmann},
  {Bochmann}, {Heinrichs}, {Meyer}, {Monecke}, {M{\"u}ller}, {Sperling},
  {{\'A}lvarez Garc{\'\i}a}, {Aparicio}, {Balaguer Jim{\'e}nez}, {Bellot
  Rubio}, {Cobos Carracosa}, {Girela}, {Hern{\'a}ndez Exp{\'o}sito}, {Herranz},
  {Labrousse}, {L{\'o}pez Jim{\'e}nez}, {Orozco Su{\'a}rez}, {Ramos},
  {Barandiar{\'a}n}, {Bastide}, {Campuzano}, {Cebollero}, {D{\'a}vila},
  {Fern{\'a}ndez-Medina}, {Garc{\'\i}a Parejo}, {Garranzo-Garc{\'\i}a},
  {Laguna}, {Mart{\'\i}n}, {Navarro}, {N{\'u}{\~n}ez Peral}, {Royo},
  {S{\'a}nchez}, {Silva-L{\'o}pez}, {Vera}, {Villanueva}, {Fourmond}, {de
  Galarreta}, {Bouzit}, {Hervier}, {Le Clec'h}, {Szwec}, {Chaigneau},
  {Buttice}, {Dominguez-Tagle}, {Philippon}, {Boumier}, {Le Cocguen},
  {Baranjuk}, {Bell}, {Berkefeld}, {Baumgartner}, {Heidecke}, {Maue}, {Nakai},
  {Scheiffelen}, {Sigwarth}, {Soltau}, {Volkmer}, {Blanco Rodr{\'\i}guez},
  {Domingo}, {Ferreres Sabater}, {Gasent Blesa}, {Rodr{\'\i}guez
  Mart{\'\i}nez}, {Osorno Caudel}, {Bosch}, {Casas}, {Carmona}, {Herms},
  {Roma}, {Alonso}, {G{\'o}mez-Sanjuan}, {Piqueras}, {Torralbo}, {Fiethe},
  {Guan}, {Lange}, {Michel}, {Bonet}, {Fahmy}, {M{\"u}ller}, \&
  {Zouganelis}}]{solanki20}
{Solanki}, S.~K., {del Toro Iniesta}, J.~C., {Woch}, J., {et~al.} 2020, \aap,
  642, A11

\bibitem[{{Solanki} {et~al.}(2013){Solanki}, {Krivova}, \& {Haigh}}]{solanki13}
{Solanki}, S.~K., {Krivova}, N.~A., \& {Haigh}, J.~D. 2013, \araa, 51, 311

\bibitem[{{Spruit} \& {Roberts}(1983)}]{spruit83}
{Spruit}, H.~C. \& {Roberts}, B. 1983, \nat, 304, 401

\bibitem[{{Unruh} {et~al.}(1999){Unruh}, {Solanki}, \& {Fligge}}]{unruh99}
{Unruh}, Y.~C., {Solanki}, S.~K., \& {Fligge}, M. 1999, \aap, 345, 635

\bibitem[{{Vieira} {et~al.}(2012){Vieira}, {Norton}, {Dudok de Wit},
  {Kretzschmar}, {Schmidt}, \& {Cheung}}]{vieira12}
{Vieira}, L.~E.~A., {Norton}, A., {Dudok de Wit}, T., {et~al.} 2012, \grl, 39,
  L16104

\bibitem[{{Yeo} {et~al.}(2015){Yeo}, {Ball}, {Krivova}, {Solanki}, {Unruh}, \&
  {Morrill}}]{yeo15}
{Yeo}, K.~L., {Ball}, W.~T., {Krivova}, N.~A., {et~al.} 2015, \jgr, 120, 6055

\bibitem[{{Yeo} {et~al.}(2014{\natexlab{a}}){Yeo}, {Krivova}, \&
  {Solanki}}]{yeo14b}
{Yeo}, K.~L., {Krivova}, N.~A., \& {Solanki}, S.~K. 2014{\natexlab{a}}, \ssr,
  186, 137

\bibitem[{{Yeo} {et~al.}(2014{\natexlab{b}}){Yeo}, {Krivova}, {Solanki}, \&
  {Glassmeier}}]{yeo14a}
{Yeo}, K.~L., {Krivova}, N.~A., {Solanki}, S.~K., \& {Glassmeier}, K.~H.
  2014{\natexlab{b}}, \aap, 570, A85

\bibitem[{{Yeo} {et~al.}(2013){Yeo}, {Solanki}, \& {Krivova}}]{yeo13}
{Yeo}, K.~L., {Solanki}, S.~K., \& {Krivova}, N.~A. 2013, \aap, 550, A95

\bibitem[{{Yeo} {et~al.}(2017){Yeo}, {Solanki}, {Norris}, {Beeck}, {Unruh}, \&
  {Krivova}}]{yeo17}
{Yeo}, K.~L., {Solanki}, S.~K., {Norris}, C.~M., {et~al.} 2017, \prl, 119,
  9.1102

\end{thebibliography}

\appendix

\section{Active region mapping}
\label{appendixar}

{We identified the ARs that emerged within P1 and were imaged by both SO/PHI and SDO/HMI. We established their boundaries as per the following procedure.}

We have 96 concurrent SO/PHI and SDO/HMI image pairs from this 10 day period (01.09.2021 to 10.09.2021). Taking each SO/PHI image pair, we created a $1024\times1024$ Boolean map and set to true the points that were classified as sunspots or faculae (Sect. \ref{tsireconstruction}), that is, harbouring magnetic activity. Let us term each contiguous cluster of magnetic points as a magnetic island. We discard the magnetic islands that do not enclose sunspots, leaving the sunspot-bearing magnetic islands (SBMIs), which we take to be the cores of active regions. To allow us to compare the Boolean maps from all the image pairs, we project each of them to a heliographic latitude versus Carrington longitude grid, namely, heliocentric coordinates. With this projection, a SBMI that is present in multiple image pairs will appear in a similar location in the corresponding Boolean maps. The Boolean maps from the 96 image pairs are combined into a single Boolean map using the logical OR operator. The result is a map that indicates the maximum extent of each SBMI. This is repeated for the SDO/HMI image pairs, producing a separate map of the SBMIs that were imaged by this instrument.

We combined the SBMI maps from the SO/PHI and SDO/HMI data sets using the logical AND operator. The result is a map of the SBMIs that appeared to both instruments, of which there are five. Taking each SBMI as the core of an active region, we expanded the boundary in every direction by 60 Mm and took the result as the final boundary of the active region. The five active regions are marked in Fig. \ref{armap}, along with their NOAA active region number. We confirmed, from visual inspection, that the boundary established here encloses each AR in {its} entirety. Each of these ARs emerged within P1, as evident from the facular and sunspot area time series {beginning near zero} (Figs. \ref{artsi}b and \ref{artsi}e). Meeting our selection criteria, these five ARs were retained for the analysis presented in Sect. \ref{section3}.

Combining the SO/PHI and SDO/HMI maps using the logical AND operator excludes SBMIs that are in one map but not the other. The discarded SBMIs correspond to NOAA active region 12860, 12865, and 12870. They do not meet our selection criteria. NOAA active region 12865, marked "C" in Fig. \ref{armap}, contained sunspots at the beginning of P1, where it was visible to SO/PHI but not to SDO/HMI. By the time it rotated onto the solar disc from SDO/HMI's perspective later in P1, the embedded sunspots had dissipated. Even though this active region was imaged by both instruments, it had emerged so much before P1 that its sunspots never appeared to SDO/HMI. As for NOAA active regions 12860 and 12870, the former was imaged only by SDO/HMI and the latter, marked 'D' in Fig. \ref{armap}, only by SO/PHI.

\end{document}